\documentclass[11pt,a4paper]{article}
\pdfoutput=1
\usepackage{jheppub}
\usepackage{mathtools}
\usepackage[all]{hypcap}
\usepackage{xcolor}
\usepackage{url}
\usepackage[utf8]{inputenc}
\usepackage[normalem]{ulem}
\allowdisplaybreaks%

\newcommand{\be}{\begin{equation}}
\newcommand{\ee}{\end{equation}}
\newcommand{\bea}{\begin{eqnarray}}
\newcommand{\eea}{\end{eqnarray}}
\newcommand{\beq}{\begin{equation}}
\newcommand{\eeq}{\end{equation}}

\renewcommand{\Re}{\operatorname{Re}}
\newcommand{\diag}{\mathop{\mathrm{diag}}\nolimits}

\begin{document}


{\raggedleft CTPU-PTC-19-17,\,PNUTP-19-A11\\}

\title{Light Higgs bosons in the general NMSSM}

\affiliation[a]{Center for Theoretical Physics of the Universe,\\
  Institute for Basic Science (IBS), Daejeon 34051, Korea}
\affiliation[b]{Department of Physics, Pusan National University, Busan 46241, Korea}

\author[a]{Kiwoon~Choi,}
\author[b]{Sang~Hui~Im,}
\author[b]{Kwang~Sik~Jeong}
\author[a]{and Chan~Beom~Park}

\emailAdd{kchoi@ibs.re.kr}
\emailAdd{imsanghui@pusan.ac.kr}
\emailAdd{ksjeong@pusan.ac.kr}
\emailAdd{cbpark@ibs.re.kr}

\abstract{
Physics beyond the Standard Model (SM) may manifest itself as small deviations from the SM predictions
for Higgs signal strengths at $125$~GeV.
Then, a plausible and interesting possibility is that the Higgs sector
is extended and at the weak scale there appears an additional Higgs
boson weakly coupled to the SM sector.
Combined with the LEP excess in $e^+e^-\to Z(h\to b\bar b)$, the diphoton excess
around $96$~GeV recently reported by CMS may suggest such a possibility.
We examine if those LEP and CMS excesses can be explained simultaneously by a singlet-like Higgs boson
in the general next-to-minimal supersymmetric Standard Model (NMSSM).
Higgs mixing in the NMSSM relies on the singlet coupling to the MSSM Higgs doublets and the higgsino
mass parameter, and thus is subject to the constraints on these supersymmetric parameters.
We find that the NMSSM can account for both the LEP and CMS excesses at 96 GeV while accommodating the observed
$125$~GeV SM-like Higgs boson.
Interestingly, the required mixing angles constrain the heavy doublet Higgs boson to be heavier than about
$500$~GeV.
We also show that the viable region of mixing parameter space is considerably modified if the higgsino
mass parameter is around the weak scale, mainly because of the Higgs coupling to photons induced by
the charged higgsinos.
}

\maketitle

\section{Introduction}

Although the Standard Model (SM) successfully describes  the observed particle physics up to energy scales
around TeV, it is clear that a more fundamental theory is needed to provide a complete description of nature.
So far the LHC has seen no clear signal for physics beyond the SM, and the discovered 125 GeV Higgs boson
has properties  compatible with the SM
predictions~\cite{ATLAS-CONF-2019-005, Sirunyan:2018koj}.
Yet an interesting possibility is that the Higgs sector is extended to include an additional light Higgs boson which is accessible to collider experiments and result in some deviations of the 125 GeV Higgs boson from the SM predictions.
The CMS has recently announced that Higgs searches in the diphoton final state show a local excess
of about $3\sigma$ at $96$~GeV~\cite{Sirunyan:2018aui}.
The results from the ATLAS  do not show a relevant excess, but are well compatible with the CMS limit~\cite{ATLAS-CONF-2018-025}.
Combined with the $2.3\sigma$ local excess observed in the LEP searches for
$e^+e^-\to Z(h \to b\bar b)$~\cite{Barate:2003sz, Schael:2006cr},
the CMS results provide a motivation  to consider the possibility that the Higgs sector involves an additional
scalar boson at $96$~GeV, which has been studied recently in
refs.~\cite{Fox:2017uwr, Richard:2017kot,
  Haisch:2017gql, Biekotter:2017xmf, Tao:2018zkx, Liu:2018xsw, Domingo:2018uim,
  Hollik:2018yek, Wang:2018vxp, Biekotter:2019kde, Cline:2019okt}.

In this paper we explore if the next-to-minimal supersymmetric SM (NMSSM) can explain
the LEP and CMS excesses around $96$~GeV while accommodating the
observed $125$~GeV Higgs boson.
The NMSSM extends the Higgs sector to include a gauge singlet scalar which generates
the higgsino mass parameter $\mu$ via its coupling $\lambda$ to the MSSM Higgs
doublets~\cite{Maniatis:2009re, Ellwanger:2009dp}.
As noticed in refs.~\cite{Choi:2012he, Choi:2013lda}, there are intriguing
relations between Higgs mixings and the model parameters
$\lambda$ and $\mu$ that hold for the general NMSSM.
Those relations are quite useful when examining how much
Higgs mixings, which determine how the neutral Higgs bosons couple to SM
particles~\cite{Ellwanger:2010nf, Jeong:2012ma, Badziak:2013bda,
 Jeong:2014xaa, Ellwanger:2015uaz, Beskidt:2017dil, Jeong:2017jqp, LiuLiJia:2019kye},
are constrained by the requirements on the model such as
radiative corrections to the Higgs masses, the perturbativity bound on $\lambda$,
and the chargino mass limit on $\mu$.
The viable region of mixing parameter space would be further constrained if one
specifies singlet self-interactions.
For instance, there are no tadpole and quadratic terms for the singlet in
the $Z_3$-symmetric NMSSM,\footnote{The
    possibility of accommodating both LEP and CMS excesses in the
    $Z_3$-symmetric NMSSM was firstly explored in
    ref.~\cite{Domingo:2018uim}. Our study considers the general NMSSM
    without any additional symmetry or matter.}
for which the mixing between the neutral singlet and
doublet Higgs bosons has a dependence on the mass of the CP-odd singlet scalar.

Our analysis is based on the relations between Higgs mixings and the model parameters,
and is performed for the general NMSSM without specifying singlet self-interactions.
We first examine if a singlet-like Higgs boson at $96$~GeV can be responsible for the LEP and CMS excesses
within the range of mixing angles allowed by the current LHC data on
the $125$~GeV Higgs boson, under the assumption that the gauginos, squarks and sleptons are heavy enough,
above TeV as indicated by the LHC searches for supersymmetry (SUSY), while the higgsinos can be significantly lighter. We then impose the constraints on $\lambda$ and $\mu$ to find the viable mixing angles.
It turns out that the general NMSSM can accommodate the SM-like 125 GeV Higgs boson compatible with the current
LHC data, and also a singlet-like 96 GeV Higgs boson explaining both the LEP and CMS excesses.
The allowed range of mixing angles is considerably modified if $\mu$ is around the weak scale
because the charged higgsinos enhance the Higgs coupling to photons.
Interestingly, if the excesses around $96$~GeV are due to the singlet-like Higgs boson,
the heavy doublet Higgs boson should be heavier than about $500$~GeV.

This paper is organized as follows.
In section~\ref{section:2}, we briefly discuss the effects of the neutral Higgs boson mixings on Higgs phenomenology
and examine the relations between the mixing angles and the NMSSM  parameters.
The region of the mixing parameter space compatible with the current LHC data on the
SM-like Higgs boson is presented in section~\ref{section:3}.
Section~\ref{section:4} is devoted to our main results, which show the mixing angles
required to explain the LEP and CMS excesses,  while satisfying the various constraints
on the NMSSM parameters.
It is also shown that the allowed mixing angles constrain the heavy Higgs boson to have a mass
in a certain range.
The final section is for the summary and comments.

\section{Higgs bosons in the general NMSSM}
\label{section:2}

In this section we describe how the neutral Higgs boson mixings depend on the NMSSM parameters, in particular,
on the singlet coupling $\lambda$ to the MSSM Higgs doublets and the higgsino mass parameter $\mu$.
Such relations should be taken into account when examining the Higgs
mixings consistent with the experimental constraints.
We also discuss the properties of Higgs bosons within the low energy effective theory constructed
by integrating out heavy superparticles under the assumption that the higgsinos can be light.
Note that our approach is applicable to a general form of NMSSM.

\subsection{Dependence of Higgs mixing on NMSSM parameters}

Taking an appropriate redefinition of superfields, one can always write the superpotential
of the general NMSSM as
\bea
W= \lambda S H_u H_d + f(S) + (\textrm{MSSM Yukawa terms}),
\eea
with a canonical K\"ahler potential.
Here $H_u$ and $H_d$ are the Higgs doublet superfields, and $S$ is the
gauge singlet superfield.
There are various types of NMSSM, depending on the form of the singlet superpotential $f(S)$.
Our subsequent discussion applies for a general form of $f(S)$, but for simplicity
we will assume no CP violation in the Higgs sector.

After the electroweak symmetry breaking, the CP-even neutral Higgs bosons
\begin{equation}
  \label{eq:rot_field}
  \begin{pmatrix}
    \hat h \\ \hat H \\ \hat s
  \end{pmatrix}
  = \sqrt2 \begin{pmatrix*}[r]
    \sin\beta & \cos\beta & 0 \\
    -\cos\beta & \sin\beta & 0 \\
    0 & 0 & 1
  \end{pmatrix*} \begin{pmatrix}
   \Re H^0_u - v\sin\beta \\
   \Re H^0_d - v\cos\beta \\
   \Re S - \langle S \rangle
  \end{pmatrix}
\end{equation}
mix with each other due to the potential terms
\begin{equation}
V_{\rm mix} = \lambda^2 |S|^2 ( |H_u|^2 +|H_d|^2 ) +
\Big( A_\lambda \lambda S H_u H_d + (\partial_S f)^* \lambda H_uH_d + {\rm h.c.} \Big),
\end{equation}
where $\langle H^0_u \rangle = v\sin\beta$ and $\langle H^0_d \rangle = v\cos\beta$ with $v=174$ GeV, and
$A_\lambda$ is the SUSY breaking trilinear coupling. With the above Higgs potential terms,
the effective Higgs $\mu$ and $B\mu$ parameters are given by
\begin{eqnarray}
\mu &=& \lambda \langle S \rangle,
\nonumber \\
B\mu &=& A_\lambda \lambda \langle S \rangle + \lambda \langle \partial_S f \rangle.
\end{eqnarray}
The supersymmetric parameters $\lambda$ and $\mu$, on which the Higgs mixing depends, are
subject to the perturbativity bound and the chargino mass bound, respectively.
Imposing the conditions for the electroweak symmetry breaking,
the mass squared matrix for $(\hat h,\hat H,\hat s)$ reads\footnote{
The value of $\hat M^2_{33}$ is determined by the singlet superpotential $f(S)$ and the associated
soft SUSY breaking terms $F(S)$ according to
\begin{equation}
\hat M^2_{33} =
 (\partial_S^2 f)^2
 + \left( \partial_S f  - \frac12\lambda v^2 \sin 2\beta \right)
 \left(\partial_S^3 f -
 \frac{ \partial_S^2 f  }{ S} \right)
 +  \frac12 \lambda v^2 \frac{A_\lambda}{ S} \sin 2\beta
 + \left(  \partial_S^2 F  - \frac{  \partial_S F }{  S } \right),
\end{equation}
where all the terms are evaluated at the vacuum.}
\begin{align}
\hat M^2_{11} &= m^2_0 + (\lambda^2 v^2 -m^2_Z ) \sin^2 2\beta,
\nonumber \\
\hat M^2_{12} &= \hat M^2_{21} = \frac{1}{2} (m^2_Z - \lambda^2 v^2) \sin 4\beta
+ \Delta m^2_{12},
\nonumber \\
\hat M^2_{13} &= \hat M^2_{31} = \lambda v (2 \mu -\Lambda \sin 2\beta),
\nonumber \\
\hat M^2_{22} &= \frac{2 |B\mu|}{\sin 2\beta} - (\lambda^2 v^2-m_Z^2) \sin^2 2\beta + \Delta m_{22}^2,
\nonumber\\
\hat M^2_{23} &= \hat M^2_{32} = \lambda v \Lambda \cos 2\beta,
\label{eq:mass_matrix}
\end{align}
with $\Lambda$ defined by
\begin{equation}
\Lambda \equiv A_\lambda + \langle \partial_S^2 f \rangle.
\end{equation}
Here $\hat M^2_{11}$, $\hat M^2_{12}$, and $\hat M^2_{22}$ include radiative
corrections, which can be sizable as arising from top and stop loops~\cite{Carena:2015moc}:
\begin{align}
  m^2_0
  &\simeq m^2_Z + \frac{3v^2 y^4_t}{4\pi^2}\left\{
    \ln\left(\frac{M^2_S}{m^2_t} \right)
    + \frac{X^2_t}{M_S^2} \left(1 - \frac{X^2_t}{12M_S^2} \right)
    \right\},
    \nonumber \\
  \Delta m^2_{12}
  &\simeq
    -\frac{3 v^2 y^4_t}{4\pi^2 \tan\beta}
    \left\{ \ln\left(\frac{M^2_S}{m^2_t} \right)
    + \frac{X_t (X_t + Y_t)}{2M_S^2} - \frac{X_t^3 Y_t}{12 M_S^4}
    \right\},
    \nonumber \\
  \Delta m_{22}^2
  &\simeq  \frac{3 v^2 y^4_t}{4\pi^2 \tan^2\beta}
    \left\{ \ln\left(\frac{M^2_S}{m^2_t} \right)
    +  \frac{X_t Y_t}{M_S^2} \left(1 - \frac{X_t Y_t}{12M_S^2} \right)
    \right\},
\label{eq:stop_corrections}
\end{align}
with $X_t= A_t -\mu \cot\beta$ and $Y_t = A_t + \mu\tan\beta$,
where $M_S=\sqrt{m_{\tilde t_1}m_{\tilde t_2}}$ is the geometric mean of the two stop mass-eigenvalues,
and $A_t$ is the SUSY breaking trilinear coupling associated with the top quark Yukawa $y_t = m_t/v$.
Note that $m_0$ corresponds to the SM-like Higgs boson mass at large $\tan\beta$
in the decoupling limit of the MSSM.
The LHC results constrain the stops to be heavier than TeV, and thus $m_0$ cannot be lower than
about $115$~GeV as long as stop mixing has
$X^2_t \lesssim 10 M_S^2$
as is  the case in the conventional mediation schemes of SUSY breaking.
The stop loop corrections maximizes $m_0$ at
$X_t = \pm \sqrt 6 M_S$,
{\em i.e.}~for maximal stop mixing.
On the other hand, the charged Higgs boson has a mass around the square-root of
$\hat M^2_{22}$,
and it should be  heavier than about $350$~GeV to avoid the experimental constraint
associated with $b\to s\gamma$~\cite{Gambino:2001ew}.

To find the mass eigenstates, one needs to diagonalize the mass squared matrix as
\bea
U\hat M^2 U^{{\rm T}} = \diag(m^2_h, \, m^2_H, \, m^2_s),
\eea
where the orthogonal mixing matrix $U$ can be parametrized as
\begin{equation}
U =
\begin{pmatrix}
    c_{\theta_1} c_{\theta_2}
    & - s_{\theta_1}
    & - c_{\theta_1} s_{\theta_2} \\
    s_{\theta_1} c_{\theta_2} c_{\theta_3} - s_{\theta_2} s_{\theta_3}
    & c_{\theta_1} c_{\theta_3}
    & - c_{\theta_2} s_{\theta_3} - s_{\theta_1} s_{\theta_2}
    c_{\theta_3} \\
    s_{\theta_1} c_{\theta_2} s_{\theta_3} + s_{\theta_2} c_{\theta_3}
    & c_{\theta_1} s_{\theta_3}
    & c_{\theta_2} c_{\theta_3} - s_{\theta_1} s_{\theta_2}
    s_{\theta_3}
  \end{pmatrix}
\end{equation}
with $s_\theta \equiv \sin\theta$ and $c_\theta \equiv \cos\theta$, for which
the angles $\theta_1$, $\theta_2$ and $\theta_3$ represent $\hat h$--$\hat H$,
$\hat s$--$\hat h$ and $\hat s$--$\hat H$ mixing, respectively.
Obviously each matrix element of $\hat M^2$ can be expressed in terms of
the mass eigenvalues $\{m_h,\,m_H,\,m_s\}$ and the mixing angles $\{\theta_1,\,\theta_2,\,\theta_3\}$.
Among such relations, the following ones are particularly relevant for our subsequent discussion:
\begin{eqnarray}
\label{important-relations}
m^2_0 &=&
m^2_h
+ U_{21}\Big( U_{21} + U_{22} \tan2\beta \Big) (m^2_H - m^2_h)
\nonumber \\
&&
-\, U_{31}\Big(U_{31}+U_{32} \tan2\beta \Big) (m^2_h- m^2_s)
- \Delta m^2_{12} \tan2\beta,
\nonumber \\
\lambda v \mu &=&
\frac{1}{2}U_{23}\Big( U_{21} + U_{22}\tan2\beta \Big) (m^2_H - m^2_h)
- \frac{1}{2}U_{33}\Big( U_{31} + U_{32}\tan2\beta \Big)(m^2_h - m^2_s),
\nonumber \\
\lambda^2v^2 &=&
m^2_Z
- \frac{2}{\sin4\beta} \left(
U_{21}U_{22}(m^2_H - m^2_h)
- U_{31}U_{32}(m^2_h - m^2_s)
- \Delta m^2_{12}  \right),
\end{eqnarray}
because the constraints on the parameters $\lambda$, $\mu$ and $m_0$ can
be converted into those on the Higgs boson masses and mixing angles,
and also vice versa.
Here one should note that $\Delta m^2_{12}$ in eq.~(\ref{important-relations}) is written
\begin{equation}
\label{Dm12}
\Delta m^2_{12}= -\frac{m^2_0 - m^2_Z}{\tan\beta} + \epsilon v^2,
\end{equation}
with $\epsilon$ given by
\begin{equation}
\epsilon = - \frac{3y^4_t}{8\pi^2}\frac{\mu}{M_S}\left(1
  +\frac{1}{\tan^2\beta} \right)
\left( \frac{X_t}{M_S} - \frac{X^3_t}{6 M_S^3} \right).
\end{equation}
The above shows that $\epsilon$ vanishes at $X_t = 0$ and
$X_t = \pm \sqrt 6 M_S$,
and therefore $\Delta m^2_{12}$ is tightly correlated with
$m_0$ near the regions of minimal and maximal stop mixing.
The correlation of the stop corrections in large regions of
parameter space has been noted in ref.~\cite{Carena:2015moc}.
It is also easy to see that $|\epsilon|$ is smaller than
about $0.1$ for $X^2_t \lesssim 10 M_S^2$.
For stop mixing with $X_t$ far from $0$ or
$\pm \sqrt 6 M_S$,
the $\epsilon$-contribution to $\Delta m^2_{12}$ can be sizable only
when $|\mu|$ is large, close to $M_S$, and $\tan\beta$ is low.
Keeping this feature in mind, we shall neglect the
$\epsilon$-contribution in our analysis unless stated otherwise.

\subsection{Effective Higgs couplings to the SM sector}

At energy scales around the electroweak scale, the properties of the Higgs bosons
can be examined within an effective theory constructed by integrating out heavy sparticles.
The LHC results on the Higgs sector and  the searches for new physics
indicate that SUSY, if exists, would be broken at a scale above TeV.
Taking this into account, we assume that all gauginos, squarks and sleptons have masses above TeV,
while the higgsinos and additional Higgs bosons can be lighter than TeV.
Then the effective lagrangian describing how the neutral Higgs bosons interact with the SM fermions and gauge bosons
is written as~\cite{Carmi:2012in}
\bea
\label{effective-action}
{\cal L}_{\rm eff} &=&
C_V^i \left(
\frac{\sqrt2m^2_W}{v}  \phi_i  W^+_\mu W^{- \mu}
+  \frac{m^2_Z}{\sqrt2 v}  \phi_i  Z_\mu Z^\mu \right)
- C_f^i \frac{m_f}{\sqrt2 v} \phi_i \bar f f
\nonumber \\
&&
+\, \Delta C_g^i \frac{\alpha_s}{12\sqrt2\pi v} \phi_i G^a_{\mu\nu} G^{a \mu\nu}
+ \Delta C_\gamma^i \frac{\alpha}{\sqrt2\pi v} \phi_i F_{\mu\nu} F^{\mu\nu},
\eea
where $(\phi_1,\,\phi_2,\,\phi_3)=(h,\,H,\,s)$ and $f$ denotes the SM fermions.

At tree level,
the Higgs couplings to massive SM particles are given by
\bea
C_V^i = U_{i1},
\quad
C_t^i =  U_{i1} - U_{i2}\cot\beta,
\quad
C_b^i = C^i_\tau = U_{1i} + U_{i2} \tan\beta.
\eea
The Higgs couplings to gluons and photons are radiatively generated, which results in
\bea
\Delta C^i_g &\simeq& A_{1/2}(\tau^i_t) C^i_t + A_{1/2}(\tau^i_b)C^i_b + \delta C^i_g,
\nonumber \\
\Delta C^i_\gamma &\simeq&
\frac{2}{9} A_{1/2}(\tau^i_t) C^i_t - \frac{7}{8} A_1(\tau^i_W)C^i_V + \delta C^i_\gamma,
\eea
where $\delta C^i_g$ and $\delta C^i_\gamma$ are additional contributions from sparticle loops, and the loop
functions are given  by
\bea
  A_{1/2} (\tau)
  &=& \frac{3}{2\tau^2} \left\{ (\tau - 1) f(\tau) + \tau \right\} ,
\nonumber \\
  A_1 (\tau)
  &=& \frac{1}{7\tau^2} \left\{ 3 (2 \tau - 1) f (\tau) + 3 \tau + 2
    \tau^2 \right\},
\eea
where $\tau^i_j \equiv m^2_\phi/(4m^2_j)$ and
\bea
  f(\tau) =
  \begin{cases}
    \arcsin^2 \sqrt{\tau} & \quad{\rm for}\,\, \tau \leq 1 \\
    -\frac{1}{4} \left( \ln \frac{1 + \sqrt{1 - \tau^{-1}}}{1 -
        \sqrt{1 - \tau^{-1}}} - i \pi\right)^2 & \quad{\rm for}\,\, \tau > 1
  \end{cases}.
\eea
For the case when the superpartners of SM particles are heavier than TeV, sparticles
give negligible contributions to $\Delta C^i_g$.
However, if $\mu$ is small, the charged higgsinos are light and can give a sizable contribution
to $\Delta C^i_\gamma$ through the following Higgs-higgsino couplings
\bea
\frac{\lambda}{\sqrt2} \sum_i U_{i3} \phi_i \tilde H^+_u \tilde H^-_d.
\eea
Under the assumption that the gauginos are significantly heavier than the higgsinos,
the Higgs coupling to photons induced by charginos can be approximated to be
\bea
\delta C^i_\gamma \simeq \frac{\lambda v}{6|\mu|}
\left( 1 + \frac{7}{30} \frac{m^2_{\phi_i}}{4|\mu|^2} \right)U_{i3}
\eea
for a Higgs boson with $m^2_{\phi_i} \ll 4|\mu|^2$.

\section{Mixing consistent with the 125~GeV Higgs boson}
\label{section:3}

For small scalar mixing, $h$ has properties close to those of the SM Higgs boson.
In this paper, we identify $h$ with the SM-like Higgs boson observed
at the LHC and examine how the scalar mixing is constrained by the
measured signal strengths.
The SM-like Higgs boson has $m_h=125$~GeV, and its couplings to the massive SM particles
are given by
\bea
    C_V^h
    = c_{\theta_1} c_{\theta_2},
\quad
    C_t^h
 = c_{\theta_1} c_{\theta_2} + s_{\theta_1} \cot\beta,
\quad
    C_b^h = C_\tau^h
    = c_{\theta_1} c_{\theta_2} - s_{\theta_1} \tan\beta,
\eea
while the couplings to gluons and photons read
\bea
\Delta C^h_g &\simeq& 0.97c_{\theta_1} c_{\theta_2}
+  (1.03\cot\beta + 0.06\tan\beta) s_{\theta_1},
\nonumber \\
\Delta C^h_\gamma &\simeq&
-0.81c_{\theta_1} c_{\theta_2}
+ 0.23 s_{\theta_1} \cot\beta - \frac{r}{6}c_{\theta_1} s_{\theta_2}.
\eea
Here $r$ is defined by
\bea \label{r_def}
r \equiv \frac{\lambda v}{|\mu|},
\eea
and measures the chargino contribution, which has been approximated by using
the fact that it is non-negligible only when $|\mu|$ is not far above the electroweak scale
for $\lambda$ below the perturbative bound, and the chargino search at LEP requires $|\mu|>104$~GeV.
One should note that
the couplings of the SM-like Higgs boson are determined by four parameters:
$\theta_1$, $\theta_2$, $r$ and $\tan\beta$.

The partial decay rates of the SM-like Higgs boson $h$ can be easily estimated  by using
the well-known decay properties of the hypothetical Higgs boson $\phi_{125}$ of the minimal SM with mass 125 GeV:
\bea
\frac{\Gamma(h\to bb)}{\Gamma(\phi_{125}\to bb)} &=&
\frac{\Gamma(h\to \tau\tau)}{\Gamma(\phi_{125} \to \tau\tau)}
= |C^h_b|^2,
\nonumber \\
\frac{\Gamma(h\to WW)}{\Gamma(\phi_{125}\to WW)} &=&
\frac{\Gamma(h\to ZZ)}{\Gamma(\phi_{125} \to ZZ)}
= |C^h_V|^2,
\nonumber \\
\frac{\Gamma(h\to gg)}{\Gamma(\phi_{125}\to gg)} &=&
\frac{|\Delta C^h_g|^2}{|\Delta C^{\phi_{125}}_g|^2},
\nonumber \\
\frac{\Gamma(h\to \gamma\gamma)}{\Gamma(\phi_{125}\to \gamma\gamma)} &=&
\frac{|\Delta C^h_\gamma|^2}{|\Delta C^{\phi_{125}}_\gamma|^2}.
\eea
Assuming that $h$ does not decay to non-SM particles, one also finds
its total decay rate to be
\bea
\frac{\Gamma_{\rm tot}(h)}{\Gamma_{\rm tot}(\phi_{125}) }  &\simeq&
0.64|C^h_b|^2 + 0.12 |C^h_t|^2 + 0.24 |C^h_V|^2,
\eea
with $\Gamma_{\rm tot}(\phi_{125})$ being the total decay width of
$\phi_{125}$.
Here we have used the branching ratios of the SM Higgs boson listed in
ref.~\cite{Heinemeyer:2013tqa}.
The production of the SM-like Higgs boson is dominated by the gluon fusion process,
and the signal strength normalized by the SM value is given by
\bea
\mu^{VV}_h
= \frac{\sigma(pp\to h)}{\sigma(pp \to \phi_{125})}
\frac{{\rm Br}(h \to VV)}{{\rm Br}(\phi_{125}\to VV)}
\simeq \frac{0.94 |\Delta C^h_g|^2|C^h_V|^2 + 0.12 |C^h_V|^4}
{0.64|C^h_b|^2 + 0.12 |C^h_t|^2 + 0.24 |C^h_V|^2},
\eea
for the inclusive $WW/ZZ$ channel, where ${\rm Br}(h\to ii)$ is the branching ratio
of the indicated mode.
For other channels, one obtains
\bea
\frac{\mu^{bb}_h}{\mu^{VV}_h} &=&
\frac{\mu^{\tau\tau}_h}{\mu^{VV}_h} = \frac{|C^h_b|^2}{|C^h_V|^2},
\nonumber \\
\frac{\mu^{\gamma\gamma}_h}{\mu^{VV}_h}
&=&
\frac{|\Delta C^h_\gamma|^2}{|\Delta C^{\phi_{125}}_\gamma|^2 |C^h_V|^2}
\simeq \frac{1.52|\Delta C^h_\gamma|^2}{|C^h_V|^2}.
\eea
It is obvious that one should have $\mu^{ii}_h=1$ if
$\theta_1=\theta_2=0$ and $r=0$.

The ATLAS collaboration has recently updated the measurements on the Higgs signal
strengths using the $13$~TeV data~\cite{ATLAS-CONF-2019-005}:\footnote{
  Both ATLAS and CMS collaborations have recently reported their
  analyses results on the Higgs coupling measurements using the LHC
  Run 2 data~\cite{ATLAS-CONF-2019-005, Sirunyan:2018koj}.
  The ATLAS analysis used the larger amount of the Run 2 data,
  up to the integrated luminosity of 80~fb$^{-1}$.
  As there is no combined global fit for the full Run 2 data yet, we
  adopted only the ATLAS result in our study.
}
\bea
\mu^{ZZ}_h = 1.13\pm 0.13,
\eea
and for the other channels
\bea
\frac{\mu^{WW}_h}{\mu^{ZZ}_h} = 0.84^{+0.18}_{-0.15},
\quad
  \frac{\mu_h^{\gamma\gamma}}{\mu_h^{ZZ}} = 0.87_{-0.12}^{+0.14}, \quad
  \frac{\mu_h^{\tau\tau}}{\mu_h^{ZZ}} = 0.86_{-0.22}^{+0.26}, \quad
  \frac{\mu_h^{bb}}{\mu_h^{ZZ}} = 0.84_{-0.27}^{+0.38}.
\eea
Note that the NMSSM leads to $\mu^{WW}_h=\mu^{ZZ}_h$, which is
within the $1\sigma$ range.

Let us now examine how severely the Higgs mixing is constrained by the LHC experimental
results on the Higgs boson at $125$~GeV.
The signal rates of $h$ are determined by two mixing angles $\theta_1$ and $\theta_2$ for
given values of $r$ and $\tan\beta$.
For instance, $\mu^{VV}_h=1$ is obtained if $\theta_1$ and $\theta_2$ satisfy\footnote{
Although we will not pursue in this paper, a large value of $\theta_1$ satisfying
\bea
\theta_1 \approx \frac{2.1-1.9 s^2_{\theta_2} }{\tan\beta}
\nonumber
\eea
can also lead to $\mu^{VV}_h=1$.
In this case, $C^h_b$ is negative and thus leads to wrong sign Yukawa couplings~\cite{Choi:2012he},
and one needs a large $\lambda$ beyond the perturbativity bound~\cite{Coyle:2018ydo}.
}
\bea
\theta_1 \approx \frac{\tan\beta}{1.4 \tan^2\beta + 1.7}\,s^2_{\theta_2},
\eea
for which the branching ratio for $h\to VV$ is suppressed compared to the case of
the SM Higgs boson, but such effect is compensated by the enhancement of production
rate via the gluon fusion process.
Note that $r$, which is relevant for the diphoton signal strength, is below about $1.2$ because
 $\lambda$ should be smaller than about 0.7 in order for the NMSSM to remain
perturbative up to the GUT scale, and $|\mu|$ should be larger than $104$~GeV to
satisfy the LEP bound on the chargino mass.

\begin{figure}[tb!]
  \begin{center}
  \includegraphics[width=0.45\textwidth]{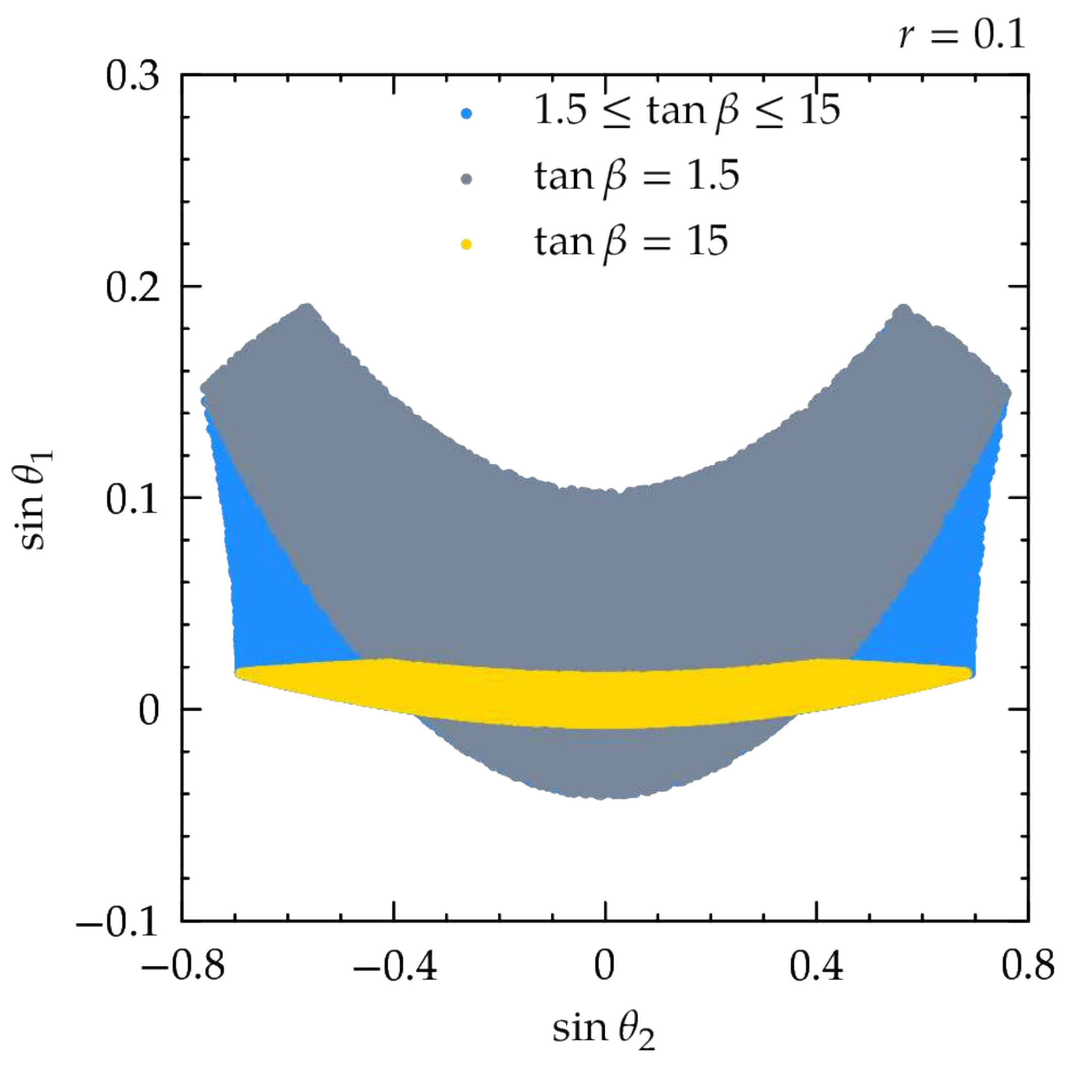}
  \includegraphics[width=0.45\textwidth]{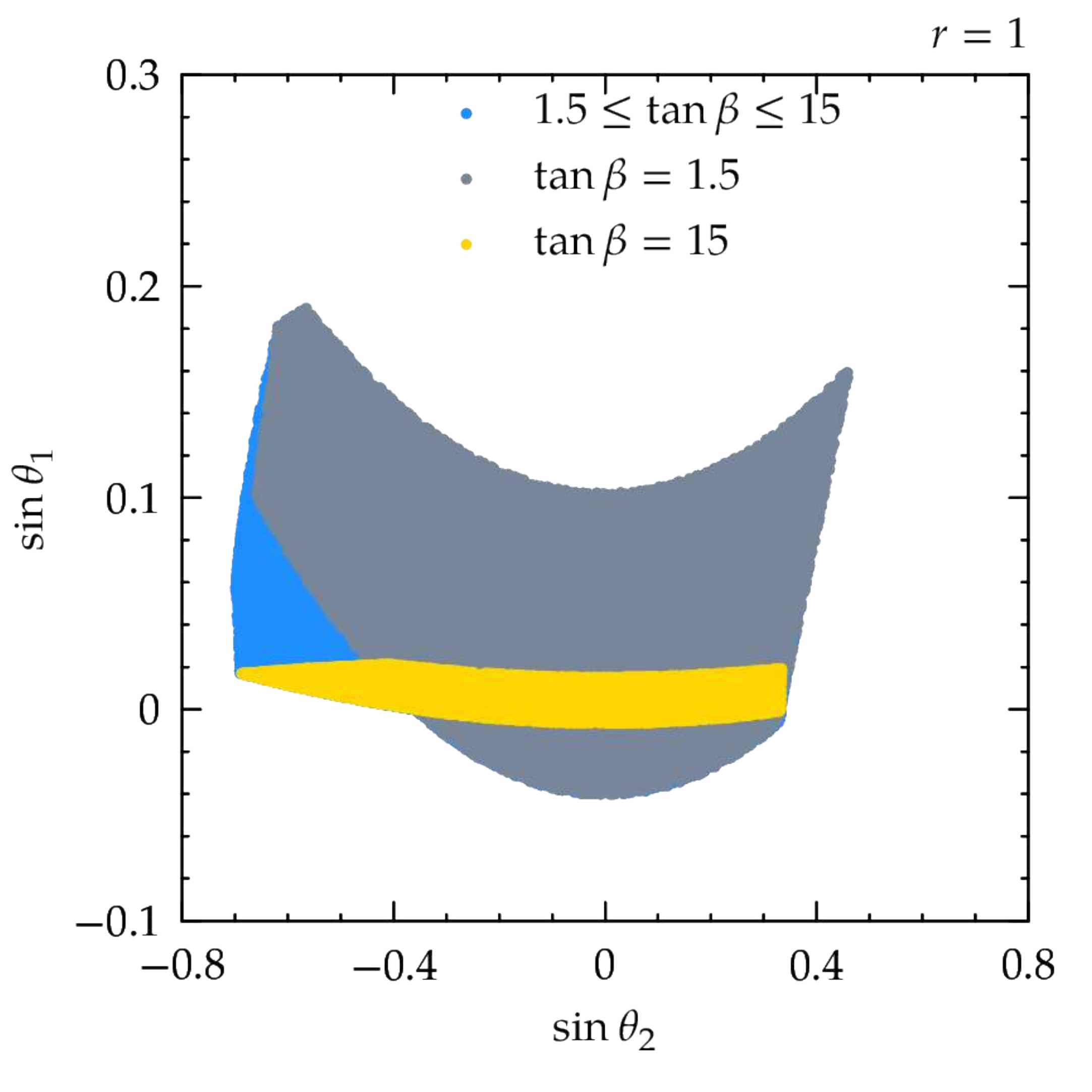}
  \end{center}
  \caption{\label{fig:sinth1_sinth2}
  Mixing angles $(\theta_1,\theta_2)$ compatible with the current LHC data on the $125$~GeV
  Higgs boson for $r=0.1$ (left) and $r=1$ (right), respectively.
  The shaded region is allowed for $1.5\leq \tan\beta \leq 15$.
  The gray and yellow colors show how the allowed region changes with $\tan\beta$.
  }
\end{figure}

Figure~\ref{fig:sinth1_sinth2} shows the $2\sigma$ range of $(\theta_1,\theta_2)$ allowed
by the current LHC data on the Higgs boson.
The left panel is for $r=0.1$, for which the Higgs coupling to photons is rarely affected by
the charged higgsinos, and the right panel is for $r=1$.
Here we have taken $1.5\leq \tan\beta \leq 15$ to see how the allowed region changes with
$\tan\beta$.
From the figure, one can see that the allowed region of $\theta_1$ gets smaller
if $\tan\beta$ increases, but a broad range of $\theta_2$ is allowed insensitively to $\tan\beta$.
This is because the Higgs coupling $C^h_b$
gets sensitive to $\theta_1$ at large $\tan\beta$ while
the couplings to the top quark and gauge bosons do not.
If $r$ is around unity, the charged higgsinos can significantly enhance the diphoton signal rate,
excluding $\theta_2$ in the range between about 0.4 and 0.8.
We can understand this feature from the fact that the charged higgsinos induce a Higgs coupling
to photons, $\delta C^h_\gamma \approx -0.17 r \theta_2$ for small mixing angles, whereas
the Higgs couplings to other SM particles only quadratically depend on $\theta_2$.

\section{\label{section:4} LEP and CMS excesses around 96~GeV}

The CMS collaboration has recently reported a local excess of
2.8$\sigma$ in the diphoton channel around $m_{\gamma\gamma} =96$~GeV~\cite{Sirunyan:2018aui}.
The signal strength amounts to
\bea
  \label{eq:CMS_excess}
\mu_{\rm CMS} =
 \frac{\sigma(pp\to \varphi)}{\sigma(pp \to \phi_{96})}
\frac{{\rm Br}(\varphi \to \gamma\gamma) }
{{\rm Br}(\phi_{96}\to\gamma\gamma)}
\simeq 0.6\pm 0.2,
\eea
where $\phi_{96}$ denotes the hypothetical SM Higgs boson with mass $96$~GeV,
and ${\rm Br}(\varphi \to \gamma\gamma)$ denotes the branching ratio for the diphoton
channel~\cite{Domingo:2018uim}.
Intriguingly, there is another 2.3$\sigma$ local excess at the similar
mass region from the Higgs searches in the $Z$-boson associated Higgs
production ($e^+ e^- \to Z \varphi$) at LEP~\cite{Barate:2003sz}.
The signal strength is~\cite{Cao:2016uwt}
\bea
  \label{eq:LEP_excess}
\mu_{\rm LEP} =
 \frac{\sigma(e^+e^-\to Z\varphi)}{\sigma(e^+e^- \to Z\phi_{96})}
\frac{{\rm Br}(\varphi \to b\bar b) }
{{\rm Br}(\phi_{96}\to b\bar b)}
\simeq
0.117 \pm 0.057.
\eea
It has long been known that the LEP excess can be explained by a light
singlet-like Higgs boson in the NMSSM\@.
At this stage, a naturally occurring question is whether this
singlet-like  Higgs boson can explain the CMS diphoton excess as well.

If both excesses were arisen due to the light singlet-like Higgs boson
$s$, the signal strengths can be expressed by the effective couplings in
eq.~(\ref{effective-action}) as follows,
\bea
\mu_{\rm CMS} &=&
\frac{\sigma(pp\to s)}{\sigma(pp \to\phi_{96})}
\frac{{\rm Br}(s \to \gamma\gamma) }
{{\rm Br}(\phi_{96}\to\gamma\gamma)}
\simeq
\frac{|\Delta C^s_g|^2 |\Delta C^s_\gamma|^2}{0.42 |C^s_b|^2 + 0.05|C^s_t|^2},
\label{mu_CMS} \\
\mu_{\rm LEP} &=&
 \frac{\sigma(e^+e^-\to Zs)}{\sigma(e^+e^- \to Z\phi_{96})}
\frac{{\rm Br}(s \to b\bar b) }
{{\rm Br}(\phi_{96}\to b\bar b)}
\simeq
\frac{|C^s_V|^2|C^s_b|^2}{0.89|C^s_b|^2+0.11|C^s_t|^2}, \label{mu_LEP}
\eea
assuming that the CP-odd singlet scalar and the singlino are heavy enough so that
$s$ decays only into the SM particles.
Here we have used \texttt{HDECAY}~\cite{Djouadi:1997yw, Djouadi:2018xqq} to calculate
the decay widths of the $96$~GeV Higgs boson with the SM couplings.
The effective couplings of $s$ are written in terms of the mixing angles as
\begin{equation}
  \begin{split}
    C_V^s
    &= s_{\theta_1} c_{\theta_2} s_{\theta_3} + s_{\theta_2}
    c_{\theta_3} , \\
    C_t^s
    &= s_{\theta_1} c_{\theta_2} s_{\theta_3} + s_{\theta_2}
    c_{\theta_3} - c_{\theta_1} s_{\theta_3} \cot\beta ,\\
    C_b^s &= C_\tau^s =
     s_{\theta_1} c_{\theta_2} s_{\theta_3} + s_{\theta_2}
    c_{\theta_3} + c_{\theta_1} s_{\theta_3} \tan\beta ,
  \end{split}
  \label{eq:coup_s_tree}
\end{equation}
and those to gluons and photons read
\begin{equation}
  \begin{split}
  \Delta C_g^s &\simeq 1.02 C_t^s - 0.08 C_b^s  ,\\
  \Delta C_\gamma^s &\simeq 0.23 C_t^s - 0.96 C_V^s
  + \frac{r}{6}U_{33},
\end{split}
\label{eq:coup_s_loop}
\end{equation}
including the contribution from the loops of charged higgsinos.

Before performing a numerical analysis, we present approximate analytic relations between mixing angles
holding if $s$ is responsible for the LEP and CMS excesses.
The LEP signal rate given in eq.~(\ref{mu_LEP}) is approximated by
\bea
\label{approximed-relation1}
\mu_{\rm LEP} \approx s^2_{\theta_2},
\eea
for small $\theta_2$,
and thus the LEP excess is explained if $s^2_{\theta_2} \sim 0.1$.
On the other hand, the ratio between the CMS and LEP signal rates is given by
\bea
\label{approximed-relation2}
\frac{\mu_{\rm CMS}}{\mu_{\rm LEP}}
\approx
\Big(
1-  k_{32} (1.1\cot\beta + 0.1\tan\beta)
\Big)^2
\times
\left(
\frac{
1
+ 0.32 k_{32} \cot\beta
-0.24s^{-1}_{\theta_2} r
}{ 1+k_{32} \tan\beta }
\right)^2,
\eea
where $k_{32}\equiv s_{\theta_3}/s_{\theta_2}$,
and it should be around $6$ to account for the LEP and CMS excesses.
Here the first factor of the right-hand side represents the Higgs production
dominated by gluon fusion, while the second one concerns the branching ratio for the diphoton mode.
Note that both effects are enhanced if $k_{32}$ is negative, and the charged higgsinos can
further increase the branching ratio into photons for $s_{\theta_2}<0$.
These features help to understand the numerical analysis given below.

Let us explain in detail how to search the viable region of mixing angles in the parameter scan.
The signal rates of $h$ and $s$ are functions of the mixing angles, a combination of $\lambda$ and $\mu$,
and $\tan\beta$:
\bea
\mu^{ii}_h &=& \mu^{ii}_h(\theta_1,\theta_2,r,\tan\beta),
\nonumber \\
\mu_{\rm LEP,\,CMS} &=& \mu_{\rm LEP,\,CMS}(\theta_1,\theta_2,\theta_3,r,\tan\beta),
\eea
for $r\equiv \lambda v/|\mu|$,
with $m_h=125$~GeV and $m_s=96$~GeV.
On the other hand, the relations~(\ref{important-relations}) exhibit how the parameters
$\lambda$, $\mu$ and $m_0$ change with the mixing angles
\bea
\lambda  &=&  \lambda(\theta_1,\theta_2,\theta_3,m_H,\tan\beta),
\nonumber \\
\mu &=& \mu(\theta_1,\theta_2,\theta_3,m_H,\tan\beta),
\nonumber \\
m_0  &=& m_0(\theta_1,\theta_2,\theta_3,m_H,\tan\beta),
\eea
where we have taken $\epsilon=0$ in eq.~(\ref{Dm12}) since  it is negligibly small in most of
the parameter space of our interest.
It is obvious that $m_H$ is determined by $\theta_{1,2,3}$ and $\tan\beta$ once we fix $r$.
The above relations allow us to analyze the viable mixing angles as follows.
We first examine the $(\theta_1, \, \theta_2)$ space to see in which region $\mu^{ii}_h$
are consistent with the current LHC data,
and then continue to check if it is further possible to explain both $\mu_{\rm LEP}$ and $\mu_{\rm CMS}$.

\begin{figure}[tb!]
  \begin{center}
  \includegraphics[width=0.45\textwidth]{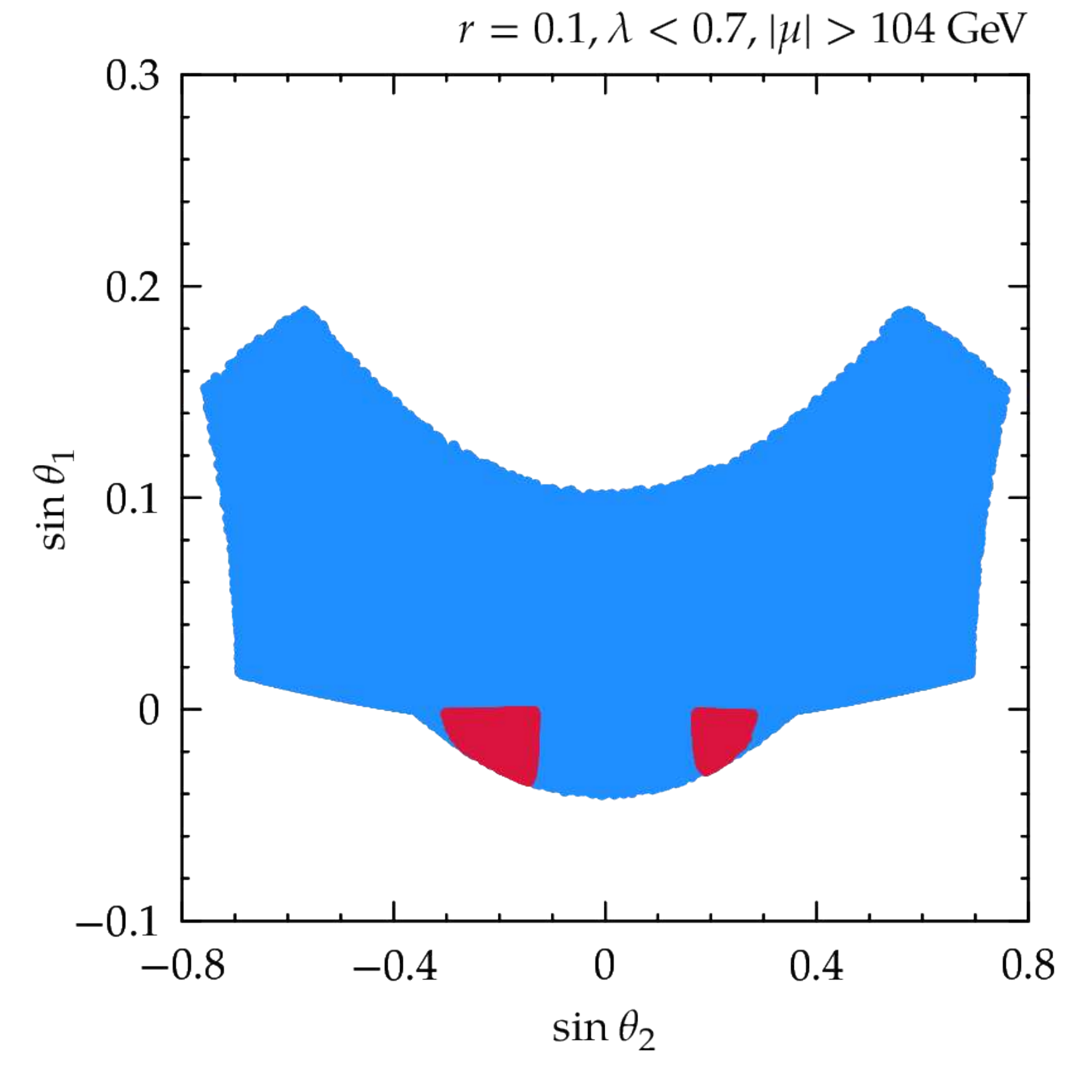}
  \includegraphics[width=0.45\textwidth]{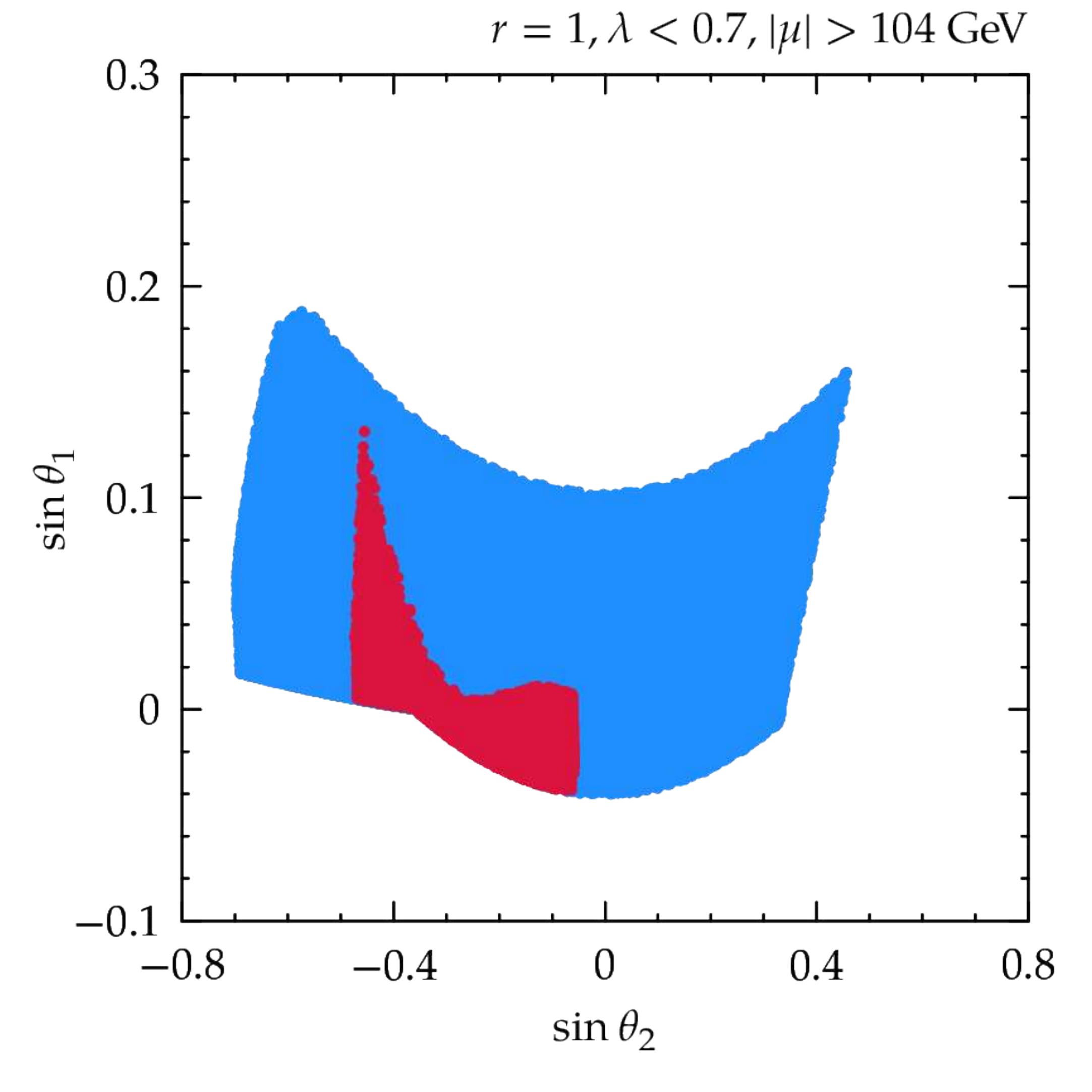}
  \end{center}
  \caption{\label{fig:sinth1_sinth2_CMS}
  Mixing angles to explain the LEP and CMS excesses for $r=0.1$ and $r=1$
  in the left and right panels,
  respectively, under the conditions $\lambda<0.7$ and $|\mu|>104$~GeV.
  The shaded region is compatible with the observed $125$~GeV Higgs boson
  for $1.5\leq \tan\beta \leq 15$ as noticed
  in figure~\ref{fig:sinth1_sinth2}.
  The singlet-like Higgs boson with mass $96$~GeV can account for
  the LEP and CMS excesses simultaneously in the red region.
  }
\end{figure}
\begin{figure}[tb!]
  \begin{center}
  \includegraphics[width=0.45\textwidth]{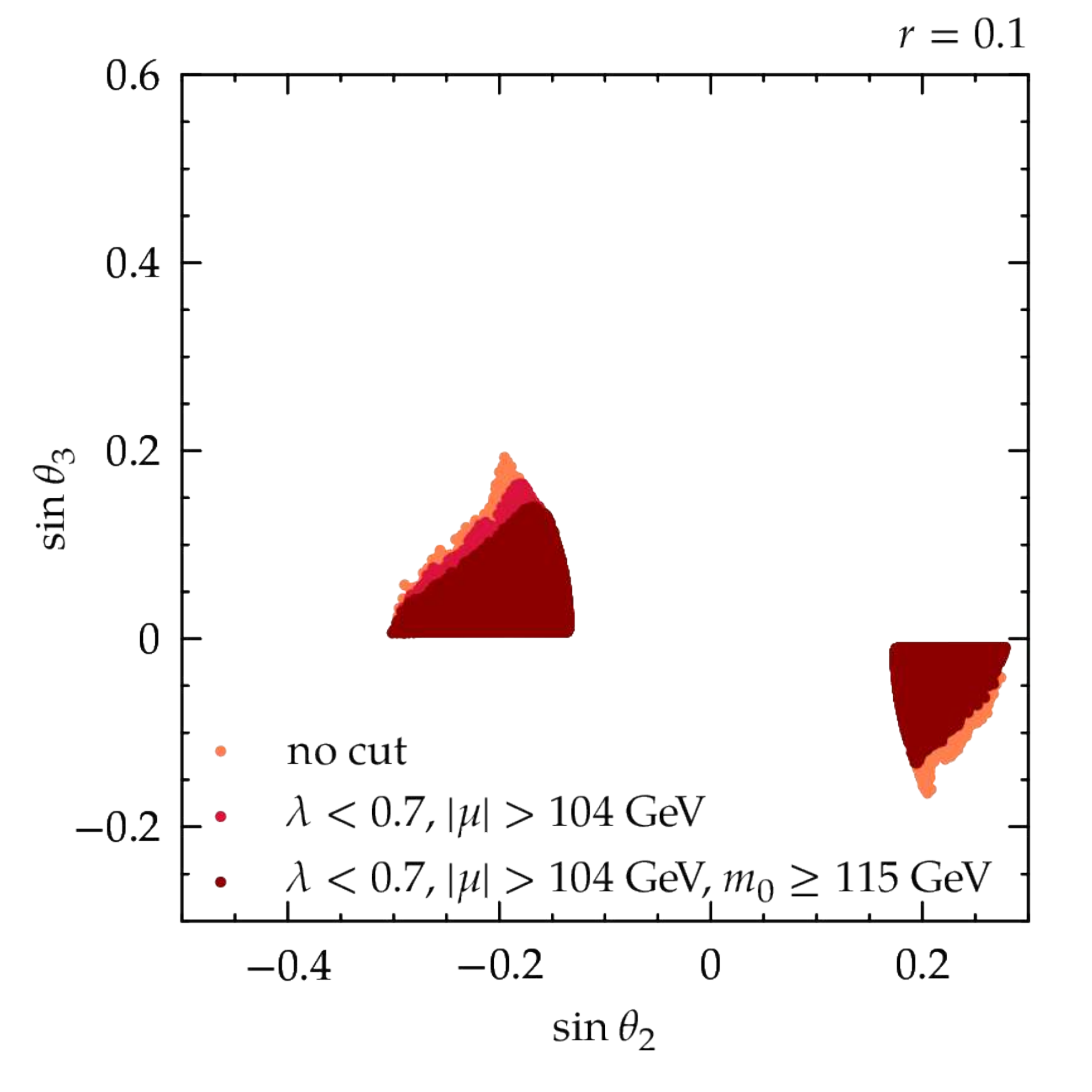}
  \includegraphics[width=0.45\textwidth]{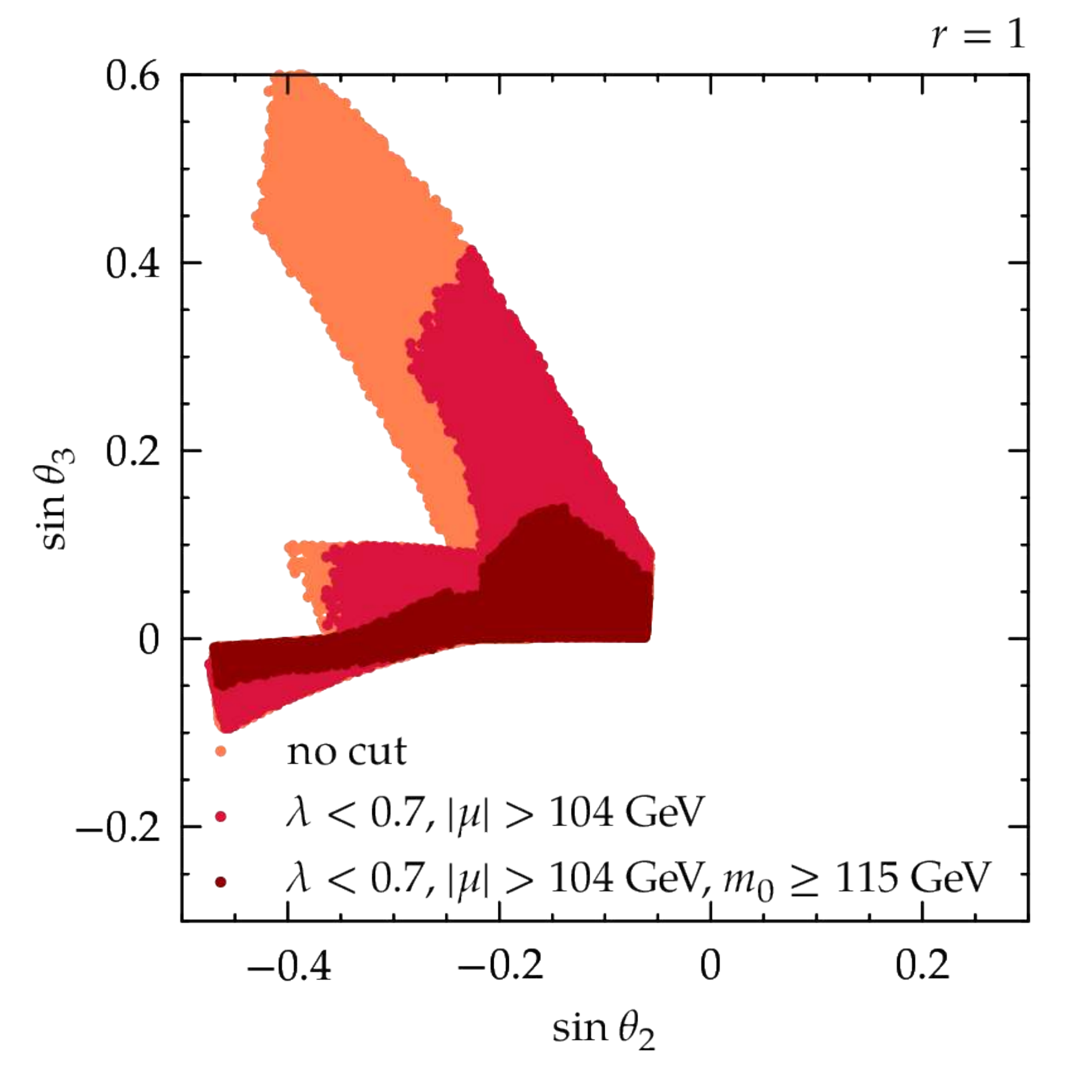}
  \end{center}
  \caption{\label{fig:sinth2_sinth3}
  Mixing angles to explain the LEP and CMS excesses for different values
  of $r$,
  continued from figure~\ref{fig:sinth1_sinth2_CMS}.
  Each color shows how the viable region of mixing parameter space is modified when the
  indicated constraint is imposed.}
\end{figure}
In figure~\ref{fig:sinth1_sinth2_CMS}, the Higgs signal rates $\mu^{ii}_h$ are consistent
with the measurements in the shaded region, and the excesses at $96$~GeV are explained in
the red shaded region.
Figure~\ref{fig:sinth2_sinth3} shows the region of $\theta_3$ for the excesses at $96$~GeV.
The next thing that one has to examine is if the viable region above is consistent also with the various
constraints on $\lambda$, $\mu$, and $m_0$.
Here we have imposed the conditions
\begin{equation}
\lambda < 0.7,\quad
|\mu|>104\,{\rm GeV},\quad
m_0\geq 115\,{\rm GeV},
\label{eq:bounds}
\end{equation}
as required by the perturbativity up to the GUT scale, the LEP limit on the chargino mass,
and the radiative contribution to the Higgs mass from stops above TeV, respectively,
provided that stop mixing is not too large as would be the case in the conventional
mediation models of SUSY breaking.\footnote{
The lower bound on $m_0$ is set by considering the numerical result that $m_0$ is around $113$~GeV
for stops at $1$~TeV in the limit of vanishing stop
mixing~\cite{Ellwanger:2004xm, Ellwanger:2005dv, Ellwanger:2006rn}.}
Then it follows $r\leq 1.1$.
Note that $r$ parameterizes the radiative effect of the charged higgsinos on Higgs decays.
As benchmark points, we have taken $r = 0.1,\,1$
for $1.5\leq \tan\beta \leq 15$.\footnote{
We have cross checked the results of our analysis for some parameter points by using
\texttt{NMSSMTools}~\cite{Ellwanger:2004xm, Ellwanger:2005dv, Ellwanger:2006rn}.
}
Each color in figure~\ref{fig:sinth2_sinth3} represents how much the
above constraints reduce the viable region.
We note that the bound on $m_0$ gets important when $\tan\beta$ is small and $r$ is
around $1$ or above.

In the parameter region for the LEP and CMS excesses,
the main effect of stop loop corrections $\Delta m_{12}^2$
is to increase (decrease)  the coupling $\lambda$
if $\Delta m_{12}^2$ is negative (positive),
as can be deduced from the last relation in eq.~(\ref{important-relations}).
This implies that the parameter space compatible with both excesses shrinks
for larger negative $\Delta m_{12}^2$ due to the perturbativity bound on $\lambda$.
In the parameter region with $m_0\geq 115$~GeV, however,
the dependence on $\Delta m_{12}^2$ becomes quite weak because $\lambda$ should be small in order
to get $m_h=125$~GeV.
On the other hand, the LEP limit on the chargino mass given in
eq.~(\ref{eq:bounds}) implies $\lambda > 0.6r$, following from
$r\equiv \lambda v/|\mu|$.
Taking this together with the perturbativity bound on $\lambda$, one
can find that $\lambda$ would be more severely constrained at larger $r$
when the stop correction $\Delta m^2_{12}$ is negative.
We have checked these features by taking analysis for nonzero values of $\epsilon$ between
$-0.05$ and $0.05$.

\begin{figure}[tb!]
  \begin{center}
  \includegraphics[width=0.45\textwidth]{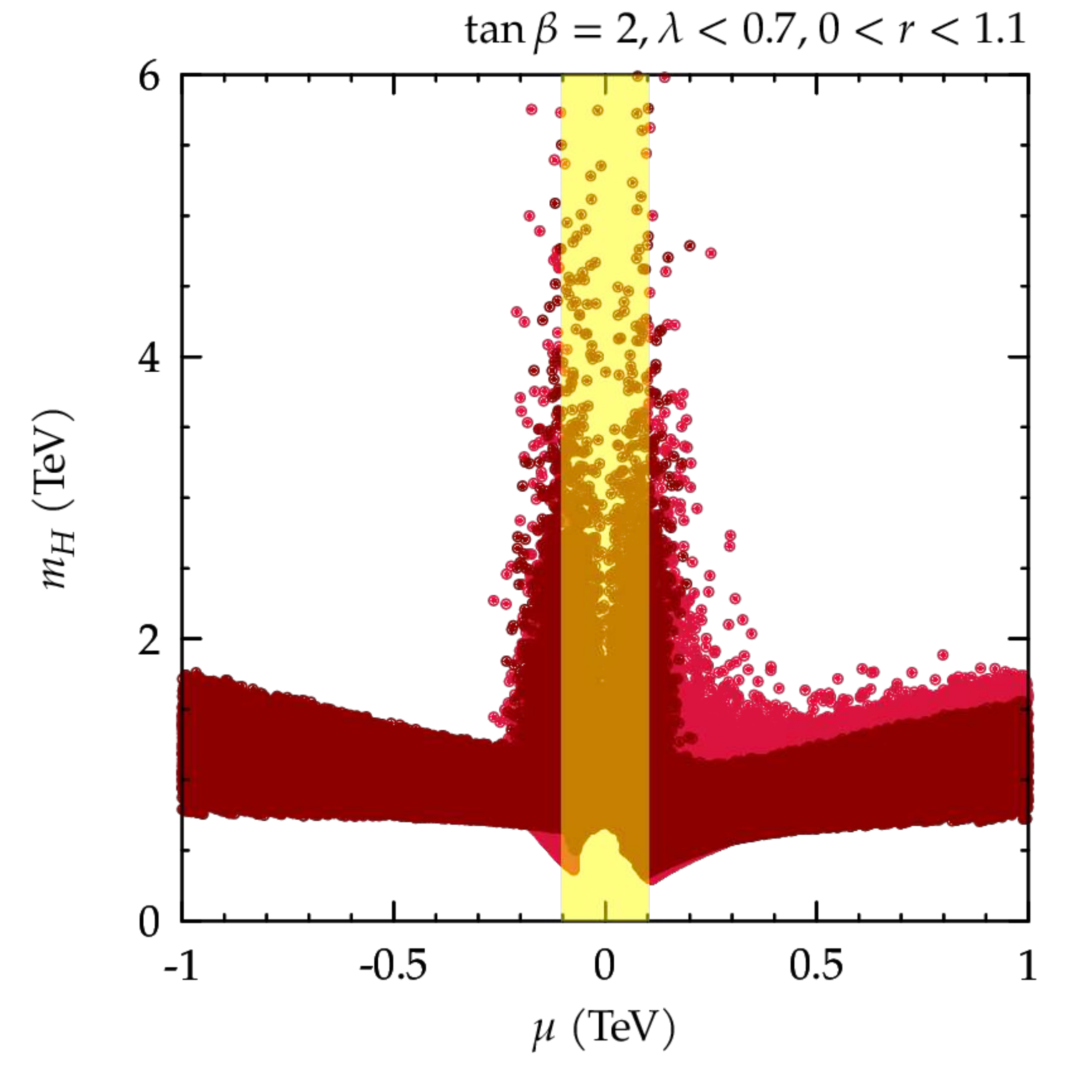}
  \includegraphics[width=0.45\textwidth]{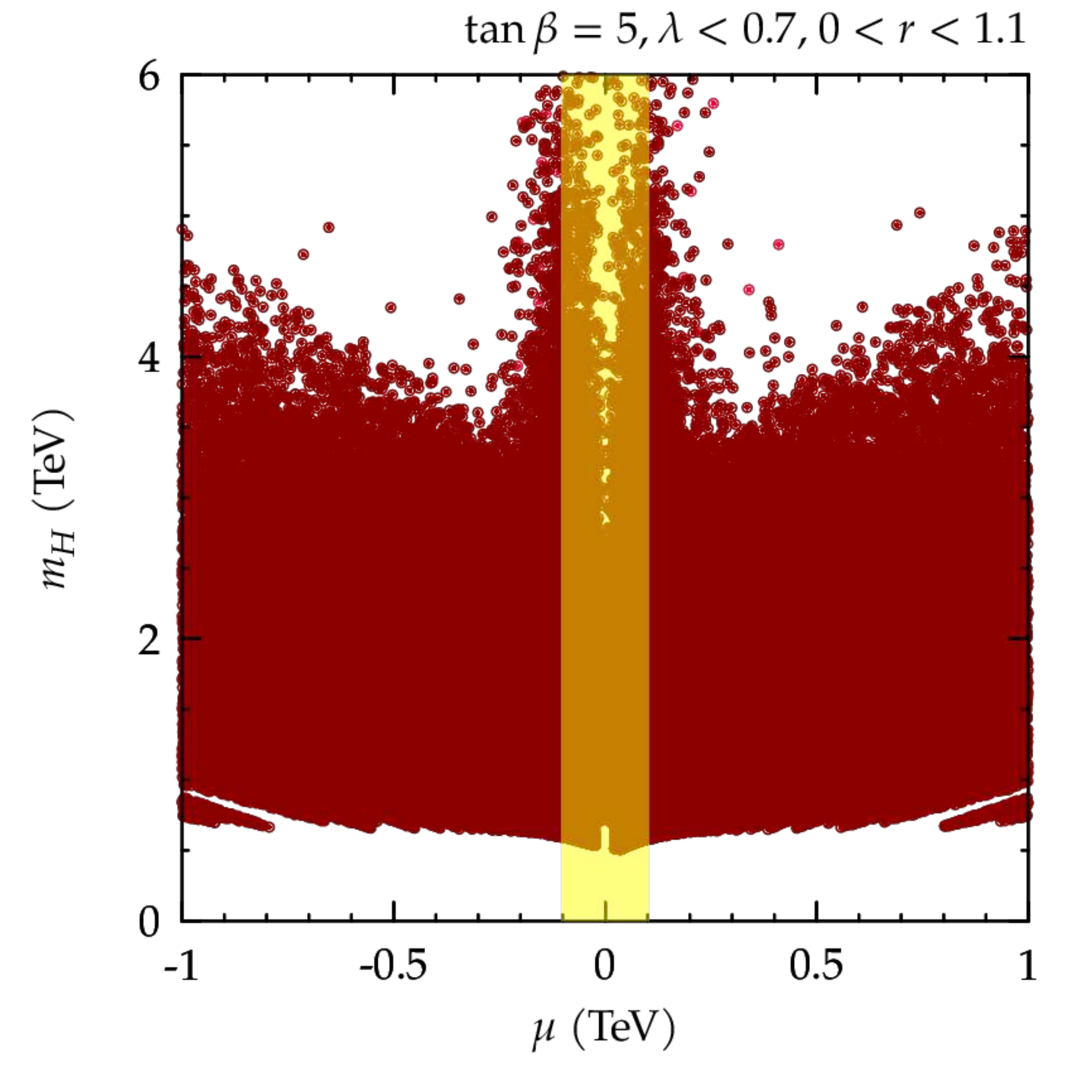}
  \end{center}
  \caption{\label{fig:mu_mh3}
  Region of $(\mu,m_H)$ compatible with the LEP and CMS excesses
  for $\tan\beta=2$ (left) and $5$ (right), respectively.
  Here we have taken $r$ smaller than $1.1$ for $\lambda$ below the
  perturbativity bound.
  In each panel, the yellow band is excluded by the LEP results on
  chargino searches
  and $m_0 < 115$~GeV in the lighter red shaded
  region. Note that the $m_0$ cut is important for $\tan\beta
  \lesssim 3$.}
\end{figure}

We close this section by pointing out that the LEP and CMS excesses can constrain
the masses of the heavy Higgs boson and higgsinos, if they are due to the singlet-like Higgs boson.
Eqs.~(\ref{important-relations}) enable us to extract the information on the region
of $\mu$ and $m_H$ compatible with the Higgs signal strengths, $\mu^{ii}_h$, $\mu_{\rm LEP}$
and $\mu_{\rm CMS}$.
Figure~\ref{fig:mu_mh3} shows the allowed region of $(\mu, m_H)$, where we have
taken $\tan\beta=2$ (left) and $5$ (right) with $0<r<1.1$.
As discussed already, the $m_0$ cut is relevant for small $\tan\beta$.
It is important to note that the CMS and LEP excesses put a lower and upper bound on $m_H$.
The lower bound turns out to be $m_H\gtrsim 500$~GeV, nearly irrespectively of the values of
$\mu$ and $\tan\beta$,
while the upper bound depends on those parameters and is found to increase
with $\tan\beta$.

\section{\label{section:5} Summary}

Extended with an additional gauge singlet scalar, the Higgs sector of the NMSSM offers a rich phenomenology
to be explored at collider experiments.
In particular, as experimentally allowed to be light, a singlet-like Higgs boson could be observable in the
searches for $e^+e^-\to Z(h\to b\bar b)$ and $pp\to h\to \gamma\gamma$ if it couples to the
SM sector via the Higgs mixing.
It is thus interesting to examine if the excesses reported by LEP and CMS in those channels can be
interpreted as signals of a singlet-like Higgs boson with mass around $96$~GeV within the NMSSM.

For the case that the gauginos, squarks and sleptons have masses above TeV, while the Higgsinos can be significantly lighter, which is perfectly consistent with the null results for SUSY searches at LHC so far, we have found that the general NMSSM can successfully
accommodate such a light singlet-like Higgs boson explaining the LEP and CMS excesses simultaneously,
as well as the $125$~GeV Higgs boson compatible with the current LHC data.
The range of mixing angles required to explain the $96$~GeV excesses can be considerably modified
if the higgsinos are around the weak scale, because the singlet-like
Higgs coupling to photons is enhanced.

To examine a viable region of mixing parameter space, it should be taken into account that
Higgs mixing is subject to various constraints on the NMSSM parameters.
We have shown that, if a singlet-like Higgs boson is responsible for the LEP and CMS excesses,
Higgs mixing is strongly constrained by the LEP bound on the charged higgsino mass and
the perturbativity bound on the singlet coupling to the Higgs doublets.
Interestingly, in the viable mixing space, the heavy doublet Higgs boson is found to be heavier than
about $500$~GeV.

The physics underlying the electroweak symmetry breaking may manifest itself as slight deviations
from the SM predictions for the Higgs signal strengths at $125$~GeV.
It is then a plausible possibility that there exist additional light Higgs bosons weakly coupled to the SM sector,
which would provide crucial information on how the Higgs sector is extended.
The excesses reported by LEP and CMS, both of which are interestingly around $96$~GeV, would
thus deserve more attention.

\acknowledgments

   We would like to thank S.~Heinemeyer for useful comments on the manuscript.
   This work was supported by IBS under the project code, IBS-R018-D1 (KC and CBP),
   and by the National Research Foundation of Korea (NRF) grant funded by the Korea
   government (MSIP) (NRF-2018R1C1B6006061) (SHI and KSJ).

\bibliography{mixing_NMSSM}
\bibliographystyle{utphys}

\end{document}